\documentclass[aps,preprintnumbers,nofootinbib,superscriptaddress]{revtex4}
\usepackage{graphicx}
\usepackage{amssymb}
\usepackage{amsmath}

\newcommand{\beq}{\begin{eqnarray}}
\newcommand{\eeq}{\end{eqnarray}}

\def\fsl#1{\setbox0=\hbox{$#1$}           
   \dimen0=\wd0                                 
   \setbox1=\hbox{/} \dimen1=\wd1               
   \ifdim\dimen0>\dimen1                        
      \rlap{\hbox to \dimen0{\hfil/\hfil}}      
      #1                                        
   \else                                        
      \rlap{\hbox to \dimen1{\hfil$#1$\hfil}}   
      /                                         
   \fi}                                         %

\begin{document}

\preprint{KEK Preprint 2011-22}

\title{Study of the conformal 
hyperscaling relation through the 
Schwinger-Dyson equation 
}

\author{Yasumichi~Aoki}
\affiliation{Kobayashi-Maskawa Institute for the Origin of Particles and the Universe, \\
Nagoya University, Nagoya 464-8602, Japan}

\author{Tatsumi~Aoyama}
\affiliation{Kobayashi-Maskawa Institute for the Origin of Particles and the Universe, \\
Nagoya University, Nagoya 464-8602, Japan}

\author{Masafumi~Kurachi}
\affiliation{Kobayashi-Maskawa Institute for the Origin of Particles and the Universe, \\
Nagoya University, Nagoya 464-8602, Japan}

\author{Toshihide~Maskawa}
\affiliation{Kobayashi-Maskawa Institute for the Origin of Particles and the Universe, \\
Nagoya University, Nagoya 464-8602, Japan}

\author{Kei-ichi~Nagai}
\affiliation{Kobayashi-Maskawa Institute for the Origin of Particles and the Universe, \\
Nagoya University, Nagoya 464-8602, Japan}

\author{Hiroshi~Ohki}
\affiliation{Kobayashi-Maskawa Institute for the Origin of Particles and the Universe, \\
Nagoya University, Nagoya 464-8602, Japan}

\author{Akihiro~Shibata}
\affiliation{Computing Research Center, High Energy Accelerator Research Organization (KEK),\\ Tsukuba 305-0801, Japan}

\author{Koichi~Yamawaki}
\affiliation{Kobayashi-Maskawa Institute for the Origin of Particles and the Universe, \\
Nagoya University, Nagoya 464-8602, Japan}

\author{Takeshi~Yamazaki}
\affiliation{Kobayashi-Maskawa Institute for the Origin of Particles and the Universe, \\
Nagoya University, Nagoya 464-8602, Japan}

\collaboration{LatKMI Collaboration}
\noaffiliation

\begin{abstract}
We study corrections to the conformal hyperscaling relation 
in the conformal window of the large $N_f$ QCD 
by using the ladder Schwinger-Dyson (SD) equation 
as a concrete dynamical model.
From the analytical expression of the solution of the 
ladder SD equation, we identify the form of the leading 
mass correction to the hyperscaling relation.
We find that the anomalous dimension, when identified through the hyperscaling relation neglecting these corrections, 
yields a value substantially lower than the one 
at the fixed point $\gamma_m^*$ for large mass region.
We further study finite-volume effects on the hyperscaling 
relation, based on the ladder SD equation 
in a finite space-time with the periodic boundary condition. 
We find that the finite-volume corrections 
on the hyperscaling relation are negligible compared with the mass correction.
The anomalous dimension, when identified through the finite-size hyperscaling relation neglecting the mass 
corrections as is often done in the lattice analyses,  yields almost the same value as that in the case of the infinite space-time neglecting the mass correction, i.e.,  
a substantially lower value than $\gamma_m^*$
for large mass. We also apply the 
finite-volume SD equation to the
chiral-symmetry-breaking phase and find that  
when the theory is close to the critical point such that  the dynamically generated mass is much smaller than 
the explicit breaking mass, the finite-size hyperscaling relation is still operative. We also suggest
a concrete form of the modification of the finite-size hyperscaling relation by including the mass correction, which may be useful to analyze the lattice data.
\end{abstract}

\maketitle

\section{Introduction}

Technicolor model \cite{tc, tcreview} has been considered as an interesting 
possibility for the dynamical origin of the electroweak symmetry breaking.
However,  it has fatal phenomenological 
difficulties  (especially with the strong suppression of 
flavor changing neutral current processes).
The problems can be solved by the walking technicolor~\cite{wtc1,wtc2} 
having 
approximate scale invariance with large mass anomalous 
dimension, $\gamma_m \simeq 1$,  which was proposed based on the ladder Schwinger-Dyson (SD) equation. 
Modern technicolor 
models often utilize 
asymptotically free gauge 
theories with an approximate 
infrared fixed point (IRFP) to achieve the walking behavior.

The SU($N$) gauge theory with a large number of massless 
fermions is one of the theories that are expected to possess such 
a property~\cite{Lane:1991qh}. 
In the case of SU($3$) gauge theory for example, the two-loop running 
coupling has an IRFP in the range of $9 \leq N_f \leq 16\ (< N_f^{\rm AF})$, 
where $N_f$ is the number of massless fermion with fundamental 
representation, and $N_f^{\rm AF}$ is the value of $N_f$ above which 
a theory loses its asymptotic freedom 
nature~\cite{Caswell:1974gg, Banks:1981nn}. 
Within this range of $N_f$, the larger the number of $N_f$ 
becomes, the smaller 
does 
the 
value of the running coupling at the IRFP. Because of this, 
it is expected that there is a critical value of flavor, $N_f^{\rm cr}$, 
below which the theory is in the confining hadronic phase with broken
chiral symmetry, while above which it is in the deconfined phase with 
unbroken chiral symmetry. An analysis based on the SD 
equation with the improved ladder approximation estimates that the 
value of $N_f^{\rm cr}$ lies between $11$ and $12$~\cite{atw}.
Therefore, 
for $12 \leq N_f \leq 16$ (often called 
 ``conformal window"), 
the theory possesses an exact IRFP, while for $9 \leq N_f \leq 11$, 
the chiral symmetry is spontaneously broken, {\it i.e.}, 
the IRFP disappears and the scale invariance is 
only approximate. In Ref.~\cite{my, Kaplan:2010zz}, this chiral phase transition at 
$N_f^{\rm cr}$ was further identified with the ``conformal phase transition" 
which was characterized by the essential singularity scaling (Miransky scaling). 

Considering the intrinsically non-perturbative nature of the problem, 
the lattice gauge theory should play  
an 
important role for the study of 
the phase structure of such theories. In addition to pioneering works 
such as Refs.~\cite{Kogut:1987ai, Iwasaki,Brown:1992fz,Damgaard:1997ut}, 
there is 
growing interest in this subject in
recent 
years~\cite{lattice}.
A straightforward way of investigating the infrared behavior of a 
given theory is to calculate the running coupling constant of the theory. 
Though it requires simulations in a wide range of parameter space 
since extensive range of the 
energy scale has to be covered to trace the 
running of the coupling by step-scaling procedure, there are many 
groups that devote their efforts to such 
a direction. 

Alternatively, infrared conformality of the theory can also be 
investigated by deforming the theory with the introduction of 
a small fermion bare mass, $m_0$, as a probe, and study the 
relation between some low-energy physical 
quantities (such as the meson masses and 
the decay constants) and $m_0$.
In Ref.~\cite{Miransky, DelDebbio:2010ze}, 
it is shown that the scaling 
relation between a low-energy quantity and $m_0$ can be expressed 
in terms of the mass anomalous dimension at the IRFP, 
$\gamma_m^\ast$.~\footnote{
This $\gamma_m^*$ is identified with the anomalous dimension $\gamma_m$  relevant 
to the walking technicolor, which is $\gamma_m$ measured at ultraviolet (UV) limit (instead of IR limit), or near the scale of 
(pseudo-) UV fixed point, usually identified with the ETC scale. See discussions below Eq.(\ref{eq:gamma}). 
} 
In the case of the mass ($M$) of a meson with certain spin and quantum 
numbers, for example, the scaling relation
(``hyperscaling relation'') is expressed as
\beq 
M \sim m_0^{1/(1+\gamma_m^\ast)}.
\label{eq:HS1}
\eeq
When one considers a theory in a finite 
space-time, the scaling relation is modified 
to the ``finite-size hyperscaling relation'' 
as follows: 
\beq
M = L^{-1} f\left( x \right), 
\label{eq:HS2}
\eeq
where, $L$ is the size of space and time, and $f$ is some function 
of  scaling variable $x$ which is defined as 
\beq
x \equiv \hat{L}\, \hat{m}_0^{1/(1+\gamma_m^\ast)}.
\eeq
Here, we introduced dimensionless quantities, 
$\hat{L}\equiv L \Lambda$ and $\hat{m}_0\equiv m_0/\Lambda$, 
where we take $\Lambda$ as the UV scale 
at which the infrared conformality terminates.
Several groups 
\cite{Fodor:2011tu, Appelquist:2011dp, 
DeGrand:2011cu, DelDebbio:2010hu}  
tried to judge whether candidate theories posses an IRFP or not 
by measuring the low-energy quantities on the lattice 
for various combination of input values of $\hat{L}$ and $\hat{m}_0$, 
then checking whether Eq.~(\ref{eq:HS2}) 
is satisfied for a certain value of $\gamma_m^\ast$.

However, a couple of questions arise here regarding use of (finite-size)
hyperscaling relation for the study of infrared conformality:
One of them is 
related to the fact that the bare fermion mass, $m_0$, which is  
introduced as a probe, itself 
necessarily 
breaks the infrared conformality 
of the original theory. How small $m_0$ has to be so that the 
hyperscaling relation is approximately satisfied? 
What is the form of correction if it is not small enough? 
When the anomalous dimension is measured for mass not so small, can it be
regarded as $\gamma_m^*$ at IR fixed point at face value? 
Another question is, when the theory in question does not have 
an IRFP (namely, in the phase where the chiral symmetry is 
spontaneously broken) in the first place, 
how and how much is the hyperscaling 
relation violated? 

The ladder SD equation, 
which is the birth place of the walking technicolor, 
is actually a concrete dynamical model 
to study such questions. In the framework of the ladder SD equation,
we know whether a given theory is infrared conformal or not, 
and also  
the value of the anomalous dimension as well. Therefore, we can 
quantitatively study the (finite-size) 
hyperscaling relation and its violation 
by using the ladder SD equation in a self-consistent manner. 
Numerical calculations can be easily 
done in a wide range of parameter space, and to a certain 
extent, even an analytical understanding can be obtained by 
investigating the solution of the ladder SD equation.

In this paper, we study the (finite-size) 
hyperscaling relation and its violation, 
based on  the ladder SD equation by taking the example of SU(3) gauge theory 
with various number of fundamental fermions (which is often 
called the large $N_f$ QCD). 

In the next section, 
from the analytical expression of the solution of the 
ladder SD 
equation, we identify the form of the leading correction to the 
hyperscaling relation.  
We find that the anomalous dimension  $\gamma_m$ for off the IRFP (with finite mass scale) is substantially smaller 
than $\gamma_m^*$, the value at IR fixed point (with vanishing mass)
for the larger mass. Our result  may shed some light on 
the value of the anomalous dimension often reported by the lattice simulations done with relatively large masses.

In Section 3, 
for the purpose of studying the finite-size 
hyperscaling relation, 
we formulate the SD equation in a finite 
space-time 
with the 
periodic boundary condition. By numerically solving it  
for various values of the
input parameters $(\hat{m}_0, \hat{L})$,  
finite-size, as well as mass deformation effects on the finite-size 
hyperscaling relation is studied in the conformal window. 
The result suggests that the correction due to the finite-size 
effect on the finite-size hyperscaling relation is negligible compared 
with that coming from large mass corrections.
Then we find that  the anomalous dimension, when identified
through the finite-size hyperscaling relation neglecting the mass corrections,
is almost the same as that obtained through the hyperscaling in the infinite space-time neglecting the mass corrections
and hence is substantially smaller than $\gamma_m^*$.
We also use 
the ladder SD equation in a finite space-time 
to study the
chiral-symmetry-breaking phase, and show how the effect of 
spontaneous breaking of the chiral symmetry affects the (finite-size) 
hyperscaling relation.  In the case of $N_f=11$ 
which is close  to the criticality so that 
the dynamically generated mass is small compared with 
the explicit mass, the finite-size hyperscaling  relation 
is still operative. 
We further suggest a concrete form of the modification 
of the finite-size hyperscaling relation taking account of the
mass correction, which may be useful for the analysis of the lattice data.

Finally, Section 5 concludes the paper.

\section{Correction to the hyperscaling relation}

We start from the study of the hyperscaling relation in  
the infinite space-time, namely the one in Eq.~(\ref{eq:HS1}). 
In this section, from the analytical expression of the solution 
of the ladder SD equation, we identify the form of the leading correction 
to the hyperscaling relation. Here, we take the example of 
the large $N_f$ QCD in the conformal window to study infrared 
conformal theories.

\begin{figure}
  \begin{center}
    \includegraphics[scale=0.4]{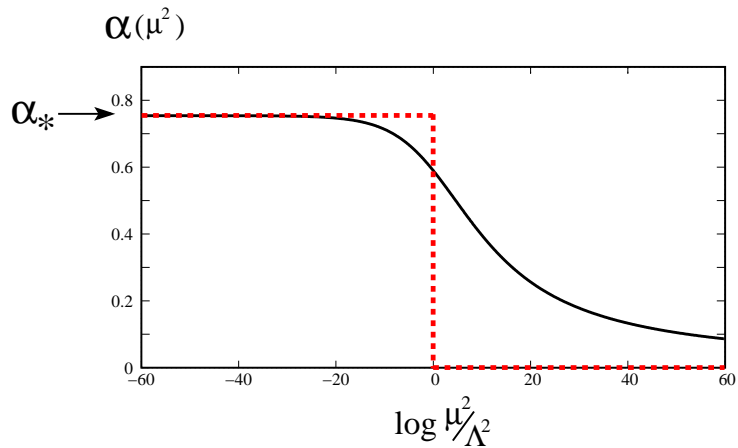}
  \end{center}
\caption{
Two-loop
running coupling (solid curve) compared with the approximate
form in Eq.~(\ref{eq:step}) (dashed line) in the case of 
SU(3) gauge theory with $12$ massless fundamental fermions.
}
\label{fig:run}
\end{figure}

The two-loop running coupling
of the large $N_f$ QCD 
is shown in Fig.~{\ref{fig:run}}.
In the figure, the two-loop running coupling (and its approximated 
form) in the case of SU(3) gauge theory with $12$ massless 
fundamental fermion is plotted as an example. 
The solid curve represents the two-loop 
running coupling which are obtained from the following RGE for 
$\alpha(\mu^2) \ \Big(= \frac{\bar{g}^2(
\mu
^2
)}{4 \pi}\,\Big)$: 
\beq
  \mu \frac{d}{d \mu} \alpha(\mu^2) 
  = \beta(\alpha(\mu^2))= -b \alpha^2(\mu^2) - c \alpha^3(\mu^2) ,
\label{eq:RGE_for_alpha}
\eeq
where
\beq
  b = \frac{1}{6 \pi} \left( 11 N_c - 2 N_f \right),\quad
  c = \frac{1}{24 \pi^2} \left( 34 N_c^2 - 10 N_c N_f
      - 3 \frac{N_c^2 - 1}{N_c} N_f \right).
      \label{b,c}
\eeq	
$\alpha_\ast$ is the value of the running coupling at the IRFP 
which is determined as
\beq
  \alpha_\ast = -  \frac{\ b\ }{\ c\ }. 
\label{eq:alpha_IR}
\eeq
Values of $\alpha_\ast$ in the case of SU(3) gauge theories 
with various number of fundamental fermion (in the conformal 
window) are shown in Table~\ref{tab:gamma}.
\begin{table}[t]
\begin{center}
\begin{tabular}{|c||c|c|c|c|c|}
\hline
\ $N_f$ \ & 12 & 13 & 14 & 15 & 16 \\
\hline
\hline
$\alpha_\ast$ & \ 0.75 \ & \ 0.47 \ & \ 0.28 \ & \ 0.14 \ & \ 0.042 \ \\
\hline
$\gamma_m^\ast$ & \ 0.80 \ & \ 0.36 \ & \ 0.20 \ & \ 0.095 \ & \ 0.027 \ \\
\hline
\end{tabular}
\caption{Values of $\alpha_\ast$ 
and also listed are
the corresponding 
$\gamma_m^\ast$ for SU(3) gauge theory with fundamental 
fermions in the conformal window to be given by Eq.(\ref{eq:gamma}).}
\label{tab:gamma}
\end{center}
\end{table}
$\Lambda$ which appears in Fig.~\ref{fig:run} 
is a renormalization group invariant scale, $\Lambda=\mu \, \exp\left(-\int^{\alpha(\mu)}\frac{d \alpha}{\beta(\alpha)}\right)$,
the two-loop analogue of $\Lambda_{\rm QCD}$ of the ordinary QCD, which is
taken as~\cite{atw}
\beq
  \Lambda \ \equiv\  \mu \ \exp \left[ - \frac{1}{b\  \alpha_\ast}
         \log \left( \frac{\alpha_\ast - \alpha(\mu^2)}{\alpha(\mu^2)}
         \right)
        - \frac{1}{b\  \alpha(\mu^2)}
         \right] ,\quad \alpha(\Lambda^2)  \simeq 0.78 \,\alpha_* ,
\label{eq:Lambda}
\eeq
in such a way that 
the scale $\Lambda$ plays the role of
the ``UV cutoff '' where the infrared conformality  
we are interested in terminates,  i.e., $\alpha(\mu^2) \sim {\rm const} (\simeq \alpha_*)$ for $\mu^2 < \Lambda^2$,  while $\alpha(\mu^2) \sim  1/\log(\mu^2/\Lambda^2)$ for
$\mu^2 >\Lambda^2$ as in the
usual asymptotically free theory.  Actually  in the walking technicolor, $\Lambda$ is taken to be of order of (or even larger than) the UV scale $\Lambda_{\rm ETC}$ (``ETC scale'') where the technicolor theory no longer makes sense as it stands and is actually converted 
into a more fundamental theory such as the Extended Technicolor (ETC).

In this paper 
we use the following form of the running coupling as an 
approximation of the two-loop running coupling 
(dashed line in Fig.~\ref{fig:run}) :
\beq
  \alpha(\mu^2) =
 \frac{ {\bar g}^2(\mu^2)}{4\pi}=
  \alpha_\ast\ \theta(\Lambda
  ^2
   - \mu
   ^2 ).
\label{eq:step}
\eeq
In this approximation the coupling takes the constant value
$\alpha_\ast$ (the value at the IR fixed point) below the scale
$\Lambda$ and entirely vanishes in the energy region above this scale.
Therefore, the physical picture of the large 
$N_f$ QCD with this approximation is the 
same as that in constant coupling gauge theory 
with UV cutoff $\Lambda$
which was extensively studied long time ago~\cite{Maskawa:1975hx, Fukuda:1976zb, Miransky:1984ef}. 
We note that it is possible, at least numerically, 
to solve the SD equation without this approximation for the two-loop 
running coupling. However, we adopt this simplification so that we 
can analytically study the solution of the SD equation to a certain extent.

\subsection{SD equation}

Let us first write down the SD equation for $SU(N_c)$ gauge theory 
with fundamental fermions:
\beq
  i S_F^{-1}(p) \ =\  \fsl{p} - m_0 + \int \frac{d^4 k}{i (2 \pi)^4}\ 
  C_2\, 
  \bar{g}^2((p-k)^2)\ \frac{1}{(p-k)^2}
  \left(g_{\mu \nu} - \frac{(p-k)_\mu(p-k)_\nu}{(p-k)^2} \right)
  \gamma^\mu \ i S_F(k) \ \gamma^\nu.
\label{eq:SDeq}
\eeq
Here,  $\ i S_{F}^{-1} \equiv A(p^2) \fsl{p} - B(p^2) $ is the full fermion propagator, 
$\ C_2 = \frac{N^2_c - 1}{2 N_c}$ is the quadratic Casimir,
and $\ \bar{g}((p-q)^2)$ is the running coupling constant.
In the above expression, we took the Landau gauge, and 
adopted the improved ladder approximation, 
in which the full gauge boson propagator is replaced by
the bare one and 
the full vertex function is replaced by a simple $\gamma^\mu$-type 
vertex with the running coupling constant associated with it. 
From this equation, we obtain 
the following two independent equations:
\beq
  A(p^2) &=& 1 + \int\frac{d^4k}{(2\pi)^4}\  
  \frac{C_2\,  \bar{g}^2((p-k)^2)
  }{k^2A(k^2)^2+B(k^2)^2} 
  A(k^2)
  \left[\frac{(p\cdot k)}{p^2(p-k)^2}
  +2\frac{\left\{p\cdot (p-k)\right\}\left\{k\cdot (p-k)\right\}}{p^2(p-k)^4}
  \right], 
  \label{eq:infSD1}\\
  B(p^2) &=& m_0 + \int\frac{d^4k}{(2\pi)^4}\  
    \frac{3\,C_2\,  \bar{g}^2((p-k)^2)
    }{k^2A(k^2)^2+B(k^2)^2}\ 
    \frac{B(k^2)}{(p-k)^2}.
    \label{eq:infSD2}
\eeq
These are coupled equations for $A(p^2)$ and $B(p^2)$ written 
in Euclidean momentum space. (Note that we have dropped 
the subscript ``$E$" for Euclidean momentum variables.)
  
To further simplify the SD equation, 
we adopt a simplified form for the argument of the running 
coupling, ${\bar g}^2 ((p-k)^2)$: 
we take it to be a function of only $p^2$ and $k^2$ 
instead of $(p-k)^2$.  
With this simplification, 
it becomes possible to carry out the 
angular integration (in the momentum space), then the SD equation 
becomes an equation for a single variable $x\equiv p_E^2$. 
Also, in this case, $A(x)=1$ is obtained from Eq.~(\ref{eq:infSD1}).  
Therefore, the SD equation 
becomes a single integral equation for the mass function 
$\Sigma(x) (\equiv B(x)/A(x) = B(x))$. 
 
For the analytical study we adopt a practically simple ansatz 
for the running coupling:~\cite{Miransky:1983vj}
\beq
{\bar g}^2 ((p-k)^2) \Rightarrow
{\bar g}^2 (max\{p^2,k^2\})
\,.
\label{ansatz1}
\eeq
Then the ladder SD equation with Eq.(\ref{eq:step}) reads:
\beq
\Sigma(x) = m_0  + \alpha_* \frac{3C_2}{4\pi} \int_0^{\Lambda^2} dy  \, 
 \frac{1}{max\{x,y\}} \,
\,\frac{\Sigma(y)}{y+\Sigma^2(y)}  \,,
\label{SD1}
\eeq
which can readily be converted into 
an equivalent 
(nonlinear) differential equation with boundary conditions~\cite{Fukuda:1976zb}:
\beq
\left( x \Sigma(x) \right)'' 
+ \alpha_\ast \frac{3 C_2}{4 \pi} \frac{\Sigma(x)}{x + \Sigma(x)^2} &=& 0, 
\label{eq:diffSDo}\\
\lim_{x\rightarrow 0} x^2 \Sigma(x)' &=& 0,
\label{eq:IRBC}\\
\left. \left( x \Sigma(x) \right)'\right|_{x=\Lambda^2} &=& m_0.
\label{eq:UVBC}
\eeq
We 
may further 
simplify 
Eq.~(\ref{eq:diffSDo}) by replacing 
$\Sigma(x)$ in the denominator of the second term in the LHS 
by a constant, $m_P$, which is 
customarily defined by :
\beq
m_P \equiv \Sigma(x=m_P^2).
\label{eq:m_P}
\eeq
Then the SD equation reads:~\cite{Fomin:1984tv} 
\beq
\left( x \Sigma(x) \right)'' 
+ \alpha_\ast \frac{3 C_2}{4 \pi} \frac{\Sigma(x)}{x + m_P 
^2} &=& 0.
\label{eq:diffSD}
\eeq
We should note that it is known that the solution obtained 
from this linearized equation well approximates that 
obtained (by numerical calculation) from the equation without linearization.
Later we shall show that particularly for the anomalous dimension there is a remarkable agreement 
between the analytical result obtained from the asymptotic solution 
of the linearized equation Eq.(\ref{eq:diffSD})
and the numerical one from the full nonlinear integral SD equation 
under a slightly different ansatz for the argument of the 
running coupling (to be mentioned later). 

$m_P$ defined in Eq.~(\ref{eq:m_P}) is often called the 
``pole mass", though of course $m_P$ is 
not the real pole mass (remember that $x$ is the Euclidean momentum 
square). Still, $m_P$ is useful quantity for the investigation of the 
hyperscaling relation since it is known, from the study 
with the Bethe-Salpeter equation~\cite{Harada:2003dc}, 
that  $m_P$ is proportional to meson masses. 
Therefore, in this paper, we use $m_P$ which is obtained from 
the solution of the SD equation as low-energy physical quantity 
which appears in the hyperscaling relation.

\subsection{Leading correction to the hyperscaling relation}
\label{sec:IIB}
Before we proceed to the investigation of the solution of the SD 
equation to derive the relation between $m_P$ and $m_0$, 
we note that it is known~\cite{Maskawa:1975hx} 
that there is a critical value of $\alpha_\ast$, $\alpha_{\rm cr} (\ne 0)$, such that 
spontaneous symmetry breaking solution ($\Sigma(x) \ne 0$) which satisfies 
Eqs.~(\ref{eq:diffSD}), (\ref{eq:IRBC}) and (\ref{eq:UVBC}) 
for the chiral limit $m_0=m_0(\Lambda)\equiv 0$ does not exists for 
\beq
 \alpha_\ast \leq 
 \alpha_{\rm cr},
 \label{eq:alphacr}
\eeq
where~\cite{Fukuda:1976zb}
\beq
\alpha_{\rm cr} 
 = \frac{\pi}{3C_2} \, =\frac{\pi}{4} \,\,(N_c=3)\,.
 \label{critical}
 \eeq
 Namely, 
   a nontrivial solution ($\Sigma(x)\neq 0$) for $\alpha<\alpha_{\rm cr}$ is the explicit breaking solution which exists only for $m_0=m_0(\Lambda) \ne 0$.
Then the pole mass $m_P$ is nothing but a renormalized mass (``current mass'') $m_R$:
\beq
m_P=m_R=Z_m^{-1} \, m_0,
\label{Zm}
\eeq
where $m_R=m_R(\mu=m_P)$,
and $Z_m=Z_m\left(\frac{\mu}{\Lambda}\right)|_{\mu=m_P}$ is the mass renormalization constant.
This means that, in the chiral 
limit $m_R=0$, there is no mass gap (dynamically generated mass) $m_D=0$, and therefore the IRFP of the 
theory is exact for $\alpha_\ast < \alpha_{\rm cr}$. 
This is exactly the region in which one expects 
that the hyperscaling relation should be satisfied. Therefore, in the 
rest of this section, we concentrate on studying the SD equation 
in the region of $\alpha_\ast < \alpha_{\rm cr}$, or equivalently 
(through the relation in Eq.~(\ref{eq:alpha_IR})) 
$
N_f^{\rm AF} \ge N_f \ge N_f^{\rm cr}
$
 (conformal window), where $N_f^{\rm AF}=16.5$ and $N_f^{\rm cr}\simeq 11.9$ for $N_c=3$.

A solution of Eq.~(\ref{eq:diffSD}) which satisfies boundary 
condition Eq.~(\ref{eq:IRBC}) can be expressed in terms of 
the hypergeometric function as~\cite{Fomin:1984tv} 
\beq
\Sigma(x) \, = \, \xi\, m_P\, F\left( \frac{1+\omega}{2}, 
\frac{1-\omega}{2}, 2, -\frac{x}{m_P^2} \right), 
\eeq
where 
\beq
\omega \equiv \sqrt{1-\frac{\alpha_\ast}{\alpha_{\rm cr}}}. 
\eeq
$\xi$ is a numerical coefficient which is determined from the 
definition of $m_P$ in Eq.~(\ref{eq:m_P}):
\beq
\xi^{-1} =  
F\left( \frac{1+\omega}{2}, \frac{1-\omega}{2}, 2, -1 \right).
\eeq
In the limit of $x \gg m_P^2$, the solution can be expanded as
\beq
\Sigma(x) \, \simeq \, \xi\, m_P\, 
\left[ 
\ 
\frac{\Gamma(\omega)}{\Gamma(\frac{\omega+1}{2})\,\Gamma(\frac{\omega+3}{2}) } 
\left( \frac{x}{m_P^2} \right)^{\frac{\omega-1}{2}}
\ +\ \ 
(\omega \leftrightarrow -\omega)
\ 
\right].
\label{eq:Sigma}
\eeq

By inserting the above expression of $\Sigma(x)$ 
into the remaining boundary condition in 
Eq.~(\ref{eq:UVBC}), we obtain the following relation between 
$m_P$ and $m_0$:
\beq
m_0 \, = \, \xi\, m_P\, 
\left[ 
\ 
\frac{\Gamma(\omega)}{\Gamma(\frac{\omega+1}{2})^2 } 
\left( \frac{\Lambda^2}{m_P^2} \right)^{\frac{\omega-1}{2}}
\ +\ \ 
(\omega \leftrightarrow -\omega)
\ 
\right].
\label{eq:m0_mP-Omega}
\eeq
From this we have the mass renormalization constant $Z_m$ in Eq.(\ref{Zm}) as
\beq
Z_m \equiv \frac{m_0}{m_R}=\frac{m_0}{m_P}=\, \xi \,
\left[ 
\ 
\frac{\Gamma(\omega)}{\Gamma(\frac{\omega+1}{2})^2 } 
\left( \frac{\Lambda^2}{m_P^2} \right)^{\frac{\omega-1}{2}}
\ +\ \ 
(\omega \leftrightarrow -\omega)
\ 
\right]\, ,
\label{Z_m}
\eeq
where we note again  $m_P=m_R$ in the conformal window. 

Then we can obtain the 
mass anomalous dimension at IRFP, $\gamma_m^\ast$, as
\beq
 \gamma_m^\ast \ =
 \lim_{m_P/\Lambda \rightarrow 0} \frac{\partial \log Z_m}{\partial \log (m_P/\Lambda)} =
 \ 1-\omega \ 
\left( =\ 1 - \sqrt{1-\frac{\alpha_\ast}{\alpha_{\rm cr}}} \ \right)\, ,
\label{eq:gamma}
\eeq
where the limit is taken as $m_P\rightarrow 0$ with
$\Lambda$ fixed.
In Table~\ref{tab:gamma}, we show values of 
$\gamma_m^\ast$, which can be calculated from the above expression 
combined with Eqs.~(\ref{b,c}), (\ref{eq:alpha_IR})
and (\ref{critical}), 
 in the case of SU(3) gauge theory with 
various numbers of fundamental fermion in the conformal window.

It should be noted that this $\gamma_m^*$ at IRFP is actually the same as the anomalous dimension in the UV limit ($\Lambda\rightarrow \infty$ with $m_P$ fixed)  $\gamma_m^{(\rm UV)} \equiv 
 \lim_{\Lambda/m_P \rightarrow \infty} \frac{\partial \log Z_m}{\partial \log (m_P/\Lambda)}=1-\omega$~\cite{Bardeen:1985sm}. $\gamma_m^{(\rm UV)} $ is the quantity  relevant to the walking technicolor with $\gamma_m^{(\rm UV)}=1$ (the value at (pseudo-) UV fixed point in the broken phase $\alpha_*>\alpha_{\rm cr}$ in the chiral limit $m_R=0$: $m_P=m_D$)~\cite{wtc1}:
 The technifermion condensate $\langle \bar \psi \psi\rangle|_{\Lambda}$ at the UV scale $\Lambda=\Lambda_{\rm ETC} (>10^3 {\rm TeV})\gg \mu \,(={\cal O} (m_P)={\cal O} ({\rm TeV})$) is enhanced by  $\gamma_m^{(\rm UV)}=1$ as
$\langle \bar \psi \psi\rangle|_{\Lambda}= Z_m^{-1}\langle \bar \psi \psi\rangle|_{\mu} $ with $Z_m^{-1} = (\Lambda/\mu)^{\gamma_m^{(\rm UV)} } =(\Lambda/\mu)^1\gg 1$.

By using this $\gamma_m^*$, we can rewrite 
the expression in Eq.~(\ref{eq:m0_mP-Omega})
in terms of 
$\gamma_m^\ast$:
\beq
\frac{m_0}{\Lambda} \ = \ \xi\,  
\left[ 
\ 
\frac{\Gamma(1-\gamma_m^\ast)}{\Gamma(\frac{2-\gamma_\ast}{2})^2 } 
\left( \frac{m_P}{\Lambda} \right)^{1+\gamma_m^\ast}
+\   
\frac{\Gamma(-1+\gamma_m^\ast)}{\Gamma(\frac{\gamma_\ast}{2})^2 } 
\left( \frac{m_P}{\Lambda} \right)^{3-\gamma_m^\ast}
\ 
\right].
\label{eq:m0-mP}
\eeq
This is the expression which should be compared with the 
hyperscaling relation in Eq.~(\ref{eq:HS1}). It is obvious that 
if we drop the second term in the RHS of Eq.~(\ref{eq:m0-mP}), 
it reduces to 
the hyperscaling relation~\cite{Miransky}. Therefore, 
the second term should be identified as the leading 
correction to the hyperscaling relation.

\begin{figure}[t]
  \begin{center}
    \includegraphics[scale=0.4]{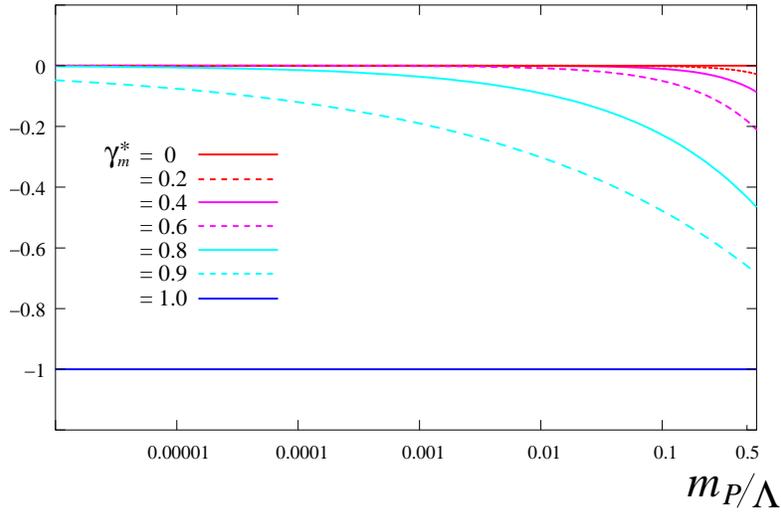}
  \end{center}
\caption{
Ratios of the second term to the first term in the RHS of 
Eq.~(\ref{eq:m0-mP}) as functions of $m_P/\Lambda$ 
for various values of $\gamma_m^\ast$. }
\label{fig:correction}
\end{figure}

To see the significance of the correction term, 
in Fig.~\ref{fig:correction}, we plot ratios of the second term 
to the first term in the RHS of Eq.~(\ref{eq:m0-mP}) as 
functions of $m_P/\Lambda$ for various values of 
$\gamma_m^\ast$ in the range of 
$0 \le \gamma_m^\ast \le 1.0$, which corresponds to 
$N_f^{\rm AF} \ge N_f \ge N_f^{\rm cr}$.
When $m_P$ is much smaller than $\Lambda$, 
(except in the case of $\gamma_m^\ast=1.0$) 
the effect of the second term is very small since in the range of 
$0 < \gamma_m^\ast  < 1.0$, the power of $(m_P/\Lambda)$ 
in the second term is always greater than that in the first term. 
This is reasonable considering the fact that small $m_P$ 
(or equivalently, small $m_0$) means small mass deformation. 
Another limit in which Eq.~(\ref{eq:m0-mP}) approximates well 
the hyperscaling is $\gamma_m^\ast \rightarrow 0$. In this limit, the 
coefficient of the second term goes to $0$ while that of the 
first term goes to $1$. Also, the power suppression of the 
second term becomes strong in this limit as well. 
However, 
we should remember that phenomenologically motivated theories 
have large $\gamma_m^\ast$.
In the limit of 
$\gamma_m^\ast \rightarrow 1$, the power of $(m_P/\Lambda)$, 
as well as coefficients of the two terms asymptote to the same values. 
Therefore, we have to take the 
second term seriously when we study 
the anomalous dimension of 
the candidate theories 
for viable walking technicolor models 
with $\gamma_m \simeq 1$ 
through the hyperscaling relation from the numerical data on the lattice.

\begin{figure}
  \begin{center}
  \vspace*{10mm}\hspace*{-14mm}
    \includegraphics[scale=0.45]{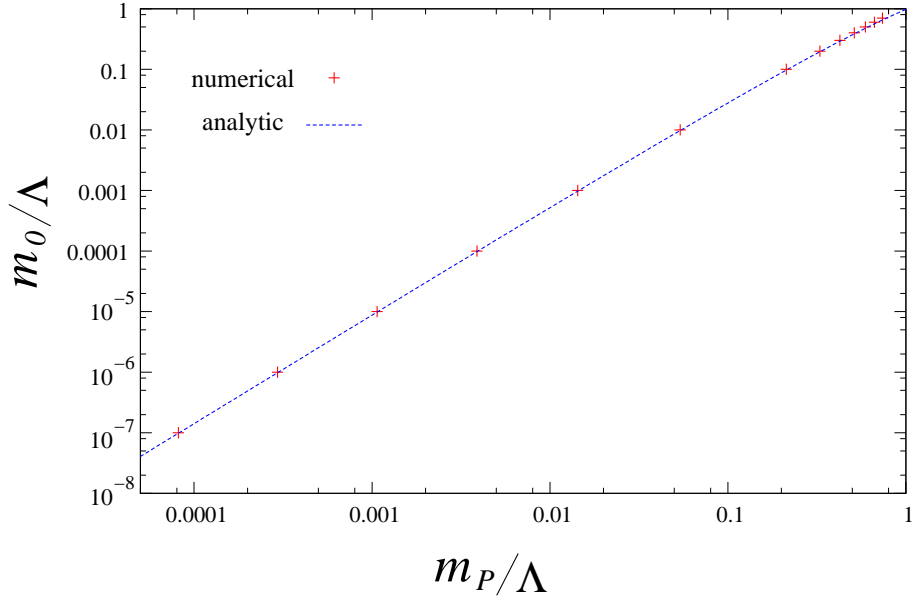}
  \end{center}
\caption{
Analytical asymptotic solution of the linearized SD equation 
vs. numerical solution of the full nonlinear one (for $N_f=12$). 
Analytic result (blue dotted line) is the plot of
Eq.~(\ref{eq:m0-mP}) which is from the asymptotic expansion of the 
linearized SD equation Eq.(\ref{eq:diffSD}) 
with the ansatz Eq.(\ref{ansatz1}), while the numerical one (denoted by red plus)
is 
that of the solution of the full nonlinear integral SD equation Eq.(\ref{SD2}) with ansatz Eq.(\ref{ansatz2}). 
}
\label{fig:LtoNL}
\end{figure}

For checking the reliability of our ansatz Eq.(\ref{ansatz1}) and 
the linearization of the differential equation, as well as the 
asymptotic expansion Eq.(\ref{eq:Sigma}),  we show 
in Fig. \ref{fig:LtoNL} the log-scale plot of 
$m_0 - m_P$  in Eq.(\ref{eq:m0-mP}) for $N_f=12$ in comparison with that obtained by directly solving 
numerically the full nonlinear 
SD equation with more natural 
(angle-averaged) ansatz:
\beq
{\bar g}^2 ((p-k)^2) \Rightarrow 
{\bar g}^2 (p^2+k^2)\, .
\label{ansatz2}
\eeq
The SD equation in this case with Eq.(\ref{eq:step}) reads:
\beq
\Sigma(x) = m_0  + \alpha_* \frac{3C_2}{4\pi} \int_0^{\Lambda^2-x} dy  \, 
 \frac{1} 
 {max\{x,y\}} \,
\,\frac{\Sigma(y)}{y+\Sigma^2(y)}  \,,
\label{SD2}
\eeq
which differs from Eq.(\ref{SD1}) by $\Lambda^2-x$ in  the UV end of the integral.
The slope in Fig.(\ref{fig:LtoNL}) corresponds to $\gamma_m +1$. The agreement is remarkable, irrespectively of the different ansatz and the additional approximations. 
%

\subsection{Effective 
anomalous dimension}
\label{ADOIRFP}
In the literature, the hyperscaling relation is often used 
as a tool to judge whether a 
theory is infrared conformal or not.
However, the importance of the corrections to the hyperscaling 
relation due to the mass deformation 
are often underestimated, 
or even completely neglected. 
This is not surprising because, in practical situations, 
it is very difficult to notice that 
data (for example, hadron mass, $M_H$, obtained from lattice 
simulations with various values of input $m_0 a$) need to be 
fitted by a function with correction term. 
Even in the situation that the correction term is not very small 
compared to the leading term, 
data could be easily fitted by a simple function of a form 
$m_0 \sim M_H^{1+\gamma}$ unless data are taken  
in a wide range of $m_0$ (or, equivalently, $M_H$).
Especially when the data are associated with, say, a few percent 
of error bars, it is very possible that one succeeds in fitting the data 
with a function of a form $m_0 \sim M_H^{1+\gamma}$. However, 
the best-fit value of $\gamma$ obtained by this fitting must be  
numerically different  
from the actual mass anomalous dimension 
at the IRFP. 
To make the difference clear, we introduce 
{\it 
effective mass anomalous dimension}, $\gamma_m^{\rm eff}$, 
which is defined as the value of $\gamma$ one obtains as a best-fit value 
when one forces to do fitting by using a fit-function which has a form 
of hyperscaling relation. 
Since the significance of the correction term is different 
for different values of $M_H$, the value of 
$\gamma_m^{\rm eff}$ should change depending on the 
range of $M_H$ one uses for fitting to obtain it.

\begin{figure}
  \begin{center}
    \includegraphics[scale=0.38]{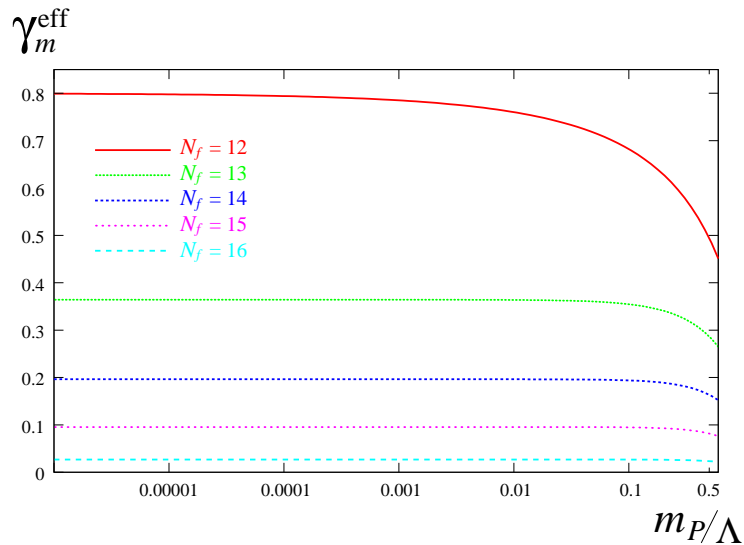}
  \end{center}
\caption{
Effective mass anomalous dimension as a function of 
$m_P/\Lambda$ for SU(3) gauge theories with 12, 
13, 14, 15 and 16 fundamental fermions.}
\label{fig:gamma-eff}
\end{figure}

\begin{figure}
  \begin{center}
    \includegraphics[scale=0.38]{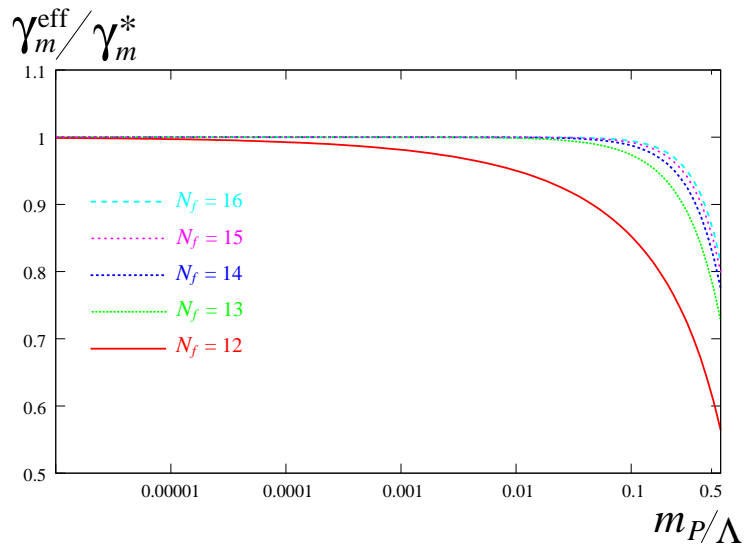}
  \end{center}
\caption{
Effective mass anomalous dimension which is normalized by 
the value of it at the IRFP  
as a function of 
$m_P/\Lambda$ for SU(3) gauge theories with 12, 
13, 14, 15 and 16 fundamental fermions.}
\label{fig:gamma-eff-norm}
\end{figure}

In the framework of the SD equation with the improved ladder 
approximation, we can 
identify 
$\gamma_m^{\rm eff}$
as $\gamma_m=\gamma_m(\mu/\Lambda)|_{\mu=m_P}$ obtained from Eq.(\ref{Z_m}) with Eq.(\ref{eq:gamma})
(or equivalently Eq.~(\ref{eq:m0-mP})): 
\beq
\gamma_m^{\rm eff} = \gamma_m &=& 
\frac{\partial \log 
Z_m }{\partial \log (m_P/\Lambda)}
\\
&=& 
\frac{\partial}{\partial \log (m_P/\Lambda)}
\log{\left( 
\xi\,  
\left[ 
\ 
\frac{\Gamma(1-\gamma_\ast)}{\Gamma(\frac{2-\gamma_\ast}{2})^2 } 
\left( \frac{m_P}{\Lambda} \right)^{
\gamma_m^\ast}
+\   
\frac{\Gamma(-1+\gamma_\ast)}{\Gamma(\frac{\gamma_\ast}{2})^2 } 
\left( \frac{m_P}{\Lambda} \right)^{
2-\gamma_m^\ast}
\ 
\right]
\right)}
\eeq
It is obvious that were it not for the second term, $\gamma_m^{\rm eff}$ would coincide with $\gamma_m^*$.
Note again that the significance of the correction term is different 
for different values of $m_P$. Therefore, the effective mass anomalous 
dimension becomes a function of $m_P$. In Fig.~\ref{fig:gamma-eff}, 
$\gamma_m^{\rm eff}$ for SU(3) gauge theories with 12, 13, 14, 
15 and 16 fundamental fermions are plotted as a function of $m_P$. 
For the purpose of making it easier to see the deviation of the 
effective mass anomalous dimension from the value at the IRFP, 
we also plot 
$\gamma_m^{\rm eff}/\gamma_m^\ast$ in 
Fig.~\ref{fig:gamma-eff-norm}.
From these figures, as we expected, we see that 
the deviation between $\gamma_m^{\rm eff}$ 
and $\gamma_m^\ast$ becomes more significant in larger 
$m_P/\Lambda$ region. We can also see that the deviation of 
the effective mass anomalous dimension is larger for 
smaller $N_f$ (or, in other words, for $N_f$ closer to 
$N_f^{\rm cr}\simeq 11.9$). This is also expected from the 
discussion below Eq.~(\ref{eq:m0-mP}) because smaller $N_f$ 
means larger $\gamma_m^\ast$, with which the correction term 
to the hyperscaling relation becomes important. As we mentioned 
earlier, phenomenologically interesting theory is 
the one with large 
mass anomalous dimension. Therefore, it is important to keep 
this effect of correction term to the hyperscaling relation in mind 
when one study such theories.

\section{Effect of the 
corrections on the finite-size hyperscaling}
So far we have concentrated on the study of the hyperscaling relation 
in the infinite 
space-time. However, since the lattice simulations 
are done in a finite 
space-time, the hyperscaling relation in 
a form of Eq.~(\ref{eq:HS2}) are used more often. Therefore, it is 
important to study the effect of correction due to mass deformation 
on the finite-size hyperscaling relation. 
For the purpose of studying the finite-size hyperscaling relation, 
we formulate the SD equation in a finite space-time with the 
periodic boundary condition. By numerically solving it  
for various values of input parameters $(\hat{m}_0, \hat{L})$,  
mass deformation effects on the finite-size hyperscaling relation 
is studied in large $N_f$ QCD.


\subsection{SD equation in a finite space-time}

To formulate the SD equation in a finite 
space-time, we start from the SD equation in the infinite 
space-time 
in Eqs.~(\ref{eq:infSD1}) and (\ref{eq:infSD2}).
To put these equations in a finite 
space-time,
all one needs to do is to replace the continuum 
momentum variables by 
the discrete ones:
\beq
 p_i \rightarrow 
  \tilde{p}_i = \frac{2\pi n_i}{L}, \ \ \ \ (n_i \in \mathbb{N}), 
\eeq
where, $p_i$ is the $i$-th component of the momentum variable.
We adopted the 
periodic boundary condition for all directions, though it is 
easy to implement the anti-periodic boundary condition. 
We also assumed that the size of all 
space-time
directions 
are the same. It is also straightforward to introduce different sizes 
for spacial and temporal directions. However, we took the same 
length for every 
direction
just for simplicity. 
$n_i$'s are integers which label 
discrete 
momentum variables. With this replacement of the momentum 
variables, 
the SD equation in Eqs.~(\ref{eq:infSD1}) and (\ref{eq:infSD2}) 
turn into the following 
form\footnote{
During the summations of $m_i$, one encounters singularities 
at $m_i = n_i$. Also, in the limit of $\tilde{B}\rightarrow 0$, 
the contribution from $(m_0, m_1, m_2, m_3)=(0, 0, 0, 0)$ 
diverges. However, these can be identified as unphysical 
artifacts considering the fact that these are integrable singularities 
in the case of infinite 
space-time. 
Therefore, in the numerical 
calculation of the SD equation, we simply drop the singular 
points from summations. (The latter singularity can also be 
avoided by adopting the anti-periodic boundary condition. We 
did the numerical calculations with the anti-periodic boundary condition 
in the temporal direction, and compared the solution with 
the one obtained from the periodic boundary condition with 
the prescription explained above. We found that the difference 
between two are negligible.) 
}:
\beq
  \tilde{A}(\tilde{p}) &=& 1 + \frac{1}{L^4} 
  \sum_{m_{0, 1, 2, 3}}
  \frac{C_2\, \bar{g}^2((\tilde{p}-\tilde{k})^2)}{\tilde{k}^2 \tilde{A}(\tilde{k})^2+\tilde{B}(\tilde{k})^2} 
  \tilde{A}(\tilde{k})\left[
  \frac{(\tilde{p}\cdot \tilde{k})}{\tilde{p}^2(\tilde{p}-\tilde{k})^2}
  +2\frac{\left\{\tilde{p}\cdot (\tilde{p}-\tilde{k})\right\}\left\{\tilde{k}\cdot (\tilde{p}-\tilde{k})\right\}}{\tilde{p}^2(\tilde{p}-\tilde{k})^4}
  \right], 
  \label{eq:finiteSD1}\\
  \tilde{B}(\tilde{p}) &=& m_0 + \frac{1}{L^4} 
  \sum_{m_{0, 1, 2, 3}} 
    \frac{3\, C_2\, \bar{g}^2((\tilde{p}-\tilde{k})^2)}{\tilde{k}^2\tilde{A}(\tilde{k})^2+\tilde{B}(\tilde{k})^2}\ 
    \frac{\tilde{B}(\tilde{k})}{(\tilde{p}-\tilde{k})^2}, 
\label{eq:finiteSD2}
\eeq
where
\beq
\tilde{p} = \frac{2\pi}{L}
\left(
  \begin{array}{c}
   n_0 \\ 
   n_1 \\ 
   n_2 \\ 
   n_3 \\ 
  \end{array}
\right),\ \ \ 
\tilde{k} = \frac{2\pi}{L}
\left(
  \begin{array}{c}
   m_0 \\ 
   m_1 \\ 
   m_2 \\ 
   m_3 \\ 
  \end{array}
\right). 
\eeq
Here, $n_i$ and $m_i$ are integers, though we should note 
that the SD equation is not defined at $\tilde{p} = 0$ since 
one of the
two independent equations is derived by requiring coefficients 
of $\fsl{p}$ in LHS and RHS are the same in Eq.~(\ref{eq:SDeq}). 
In the above expressions, $A(p^2)$ and $B(p^2)$ were replaced by 
$\tilde{A}(\tilde{p})$ and $\tilde{B}(\tilde{p})$. 
This is because they are 
no longer functions of momentum-squared since 
the rotational symmetry is broken (except some residual discrete 
rotational symmetry) due to the hyper-cubic 
shape of the finite-size space-time.
However, 
at the practical level, the effect of such rotational-symmetry 
violation is negligible 
as far as $L$ is taken large enough compared to the 
scale of relevant physics. This is true in the case 
of the
current study. We first numerically solve, by using 
iteration method, Eqs.~(\ref{eq:finiteSD1}) 
and (\ref{eq:finiteSD2}) as coupled equations for 
$\tilde{A}(\tilde{p})$ and $\tilde{B}(\tilde{p})$. Then, 
to obtain the value of $m_P$ (see Eq.~(\ref{eq:m_P})), 
we plot $\tilde{\Sigma}(\tilde{p}) \equiv 
\tilde{B}(\tilde{p})/\tilde{A}(\tilde{p})$ as a function of 
$\tilde{p}^2$. There are always multiple $\tilde{p}$'s 
which give the same value of $\tilde{p}^2$, and those do 
not necessarily give a degenerate value of $\tilde{\Sigma}$ 
unless those are related by the residual discrete rotational 
symmetry. However, we confirmed that such differences are 
negligible in the region where $m_P$ is determined.
We should also note that, when we estimate the value of $m_P$ 
from Eq.~(\ref{eq:m_P}), we used a function which is obtained 
by interpolating $\tilde{\Sigma}(\tilde{p})$ in momentum space. 
However, we never did extrapolation to the scale below 
$2\pi/L$ since there is no reliable information below 
that scale. Therefore, we obtain data only when 
the value of $m_P^2$ is greater $(2\pi/L)^2$.

\subsection{Finite-size hyperscaling and its corrections}

In this subsection, by numerically solving the finite-volume SD equation 
formulated in the previous subsection, we generate data 
of $m_P$ for various sets of input parameters 
$(L \Lambda, m_0/\Lambda)$. We take SU(3) gauge theory 
with 12 fundamental fermion as an example here. 
Then, by using those generated data, we do the analysis 
based on the finite-size hyperscaling in Eq.~(\ref{eq:HS2}). 
This is a kind of ``simulation" of the practical situation 
we often encounter when we study a theory by using 
data obtained from lattice simulations. 
An interesting point about doing hyperscaling 
analysis using data generated by the SD equation 
is that we know that the SU(3) gauge theory with 12 
fundamental fermions, in the framework of the SD equation, 
{\it is} the infrared conformal theory, and we also know the 
value of the mass anomalous dimension at the IRFP, which 
is estimated as $\gamma_m^\ast \simeq 0.80$ in this case. 
(See Table~\ref{tab:gamma}.)
Therefore, 
we clearly see how the finite-size hyperscaling is violated 
due to the effect of the mass deformation.

\begin{figure}
  \begin{center}
    \includegraphics[scale=0.59]{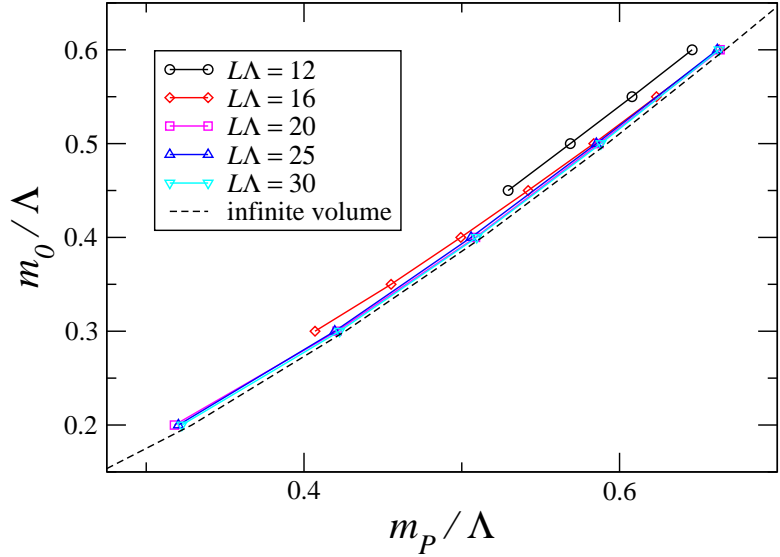}
  \end{center}
\caption{
Values of $m_P/\Lambda$ (horizontal axis) for various values of 
$m_0/\Lambda$ (vertical axis) and $L \Lambda$ 
(indicated by different symbols) 
for SU(3) gauge theory with 12 fundamental fermions.
Dashed curve is $m_P$ as a function of $m_0$ which is 
obtained from the numerical solution of the SD equation in 
the infinite space-time
 (Eqs.~(\ref{eq:infSD1}) and (\ref{eq:infSD2})).
}
\label{fig:Nf12}
\end{figure}

In Fig.~\ref{fig:Nf12}, we plot the values of 
$m_P/\Lambda$ (horizontal axis) 
for various values of $m_0/\Lambda$ (vertical axis) and $L \Lambda$ 
(indicated by different symbols). 
When we solved the finite volume SD equation, 
we adopted an angle averaged form 
(as in Eq.~(\ref{ansatz2})) 
for the argument of the running 
coupling. As we explained in the previous section, 
this is the procedure which is needed to make the angular  
integration (in momentum space) possible in the 
case of infinite-volume SD equation, and it is actually not 
needed for 
finite-volume SD equation since we numerically solve them by 
iteration without doing angular integration. However, for the purpose 
of putting finite- and infinite-volume SD equations on the same ground, 
we adopted angle averaged argument 
for finite-volume SD equation as well. 
We note that when we adopt the angle-averaged form for the 
argument of the running coupling, summations in the finite-volume 
SD equation are restricted in the range of 
$\tilde{p}^2 + \tilde{k}^2 \le \Lambda^2$.

In Fig.~\ref{fig:Nf12}, 
in each $L \Lambda$, one notices that data are plotted in the range of  
$m_P/\Lambda$ which is larger than a certain value. This lower 
limit comes from the IR cutoff effect which was explained 
at the end of the previous subsection. Dashed curve in the figure 
is $m_P$ as a function of $m_0$ which is obtained from the 
numerical solution of the SD equation in the
infinite space-time 
(Eqs.~(\ref{eq:infSD1}) and (\ref{eq:infSD2}) with the 
ansatz Eq.(\ref{ansatz2}) and Eq.(\ref{eq:step})). 
In the figure we see that 
data for $L \Lambda = 20, 25, 30$ are almost degenerate, 
and take values close to the dashed curve for 
the infinite 
space-time. 
This means that $L \Lambda = 20$ is large enough that the  
finite-size effect is negligible for the determination of $m_P$ 
in this mass range.

\begin{figure}
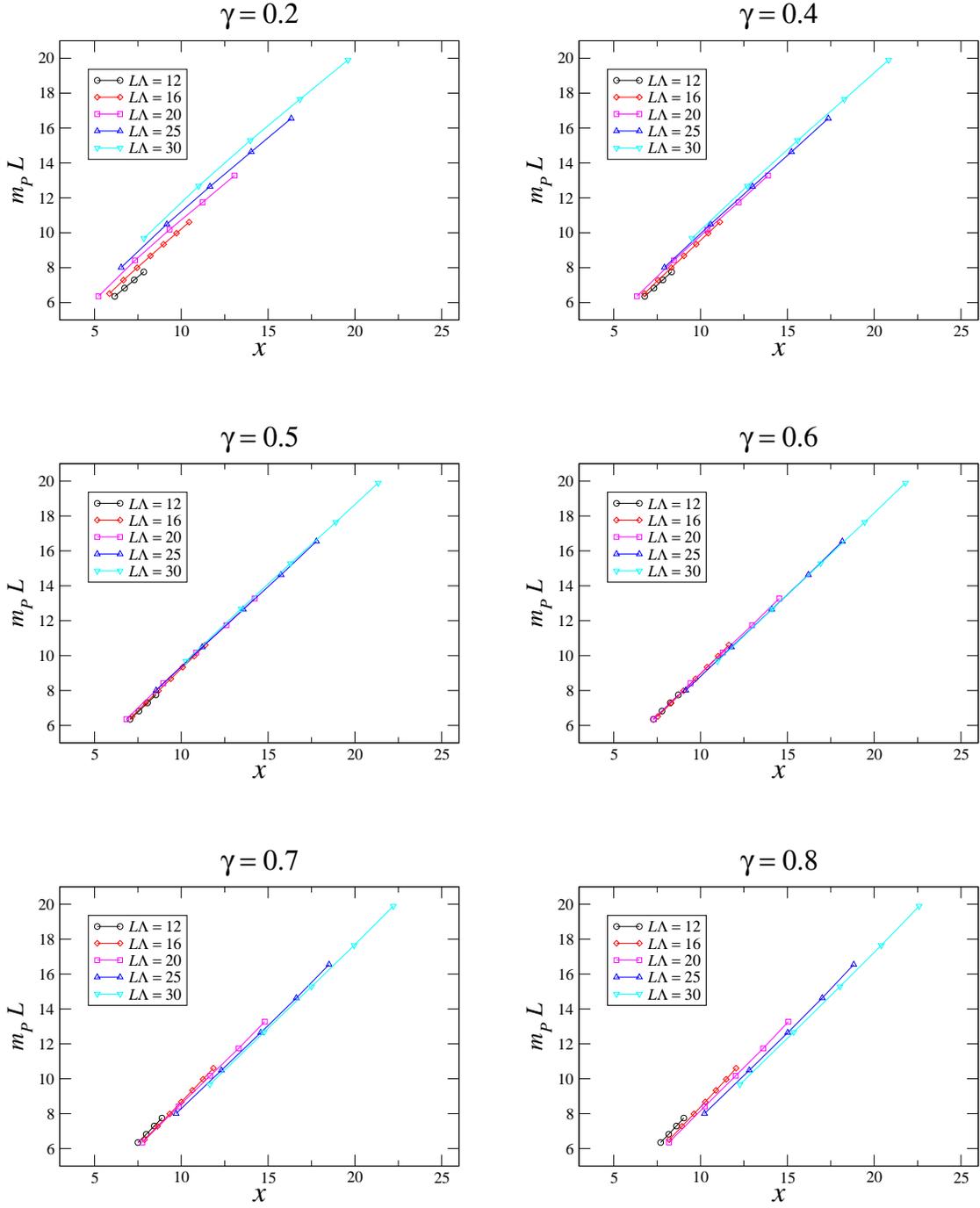

  \begin{center}
    \includegraphics[scale=0.4]{Nf12g0-2.eps}\ \ \ \ \ \ \ \ \ 
    \includegraphics[scale=0.4]{Nf12g0-4.eps}\\
\vspace*{10mm}
    \includegraphics[scale=0.4]{Nf12g0-5.eps}\ \ \ \ \ \ \ \ \ 
    \includegraphics[scale=0.4]{Nf12g0-6.eps}\\
\vspace*{10mm}
    \includegraphics[scale=0.4]{Nf12g0-7.eps}\ \ \ \ \ \ \ \ \ 
    \includegraphics[scale=0.4]{Nf12g0-8.eps}
  \end{center}
\caption{
Values of $m_P L$ obtained from the finite-volume SD equation 
as a function of 
$x\equiv L\Lambda (m_0/\Lambda)^{1/(1+\gamma)}$ for 
$\gamma = 0.2, 0.4, 0.5, 0.6, 0.7$ and $0.8$ in 
SU(3) gauge theory with 12 fundamental fermions.
Data for $L\Lambda =12, 16, 20, 25$ and $30$ are plotted 
as different symbols.
}
\label{fig:Nf12HS}
\end{figure}

Now, let us do the finite-size hyperscaling analysis by using 
data shown in Fig.~\ref{fig:Nf12}. In Fig.~\ref{fig:Nf12HS}, 
we plot the values of $m_P L$ as a function of 
$x\equiv L\Lambda (m_0/\Lambda)^{1/(1+\gamma)}$ for 
$
\gamma =
 0.2, 0.4, 0.5, 0.6, 0.7$ and $0.8$. If the theory 
is infrared conformal, and if the effect of the mass deformation 
is negligible, this kind of plot should show good alignment of 
data when 
input value of $\gamma$ is chosen to be $\gamma_m^*$, 
the value of 
mass anomalous dimension at the IRFP. Here, we know that, 
in the framework of the SD equation, the theory is infrared 
conformal, and the value of the mass anomalous dimension 
at the IRFP is $\gamma_m^\ast \simeq 0.8$. However, 
plot in Fig.~\ref{fig:Nf12HS} shows no alignment for $\gamma=0.8$, 
instead, data are well aligned for $\gamma=0.5$ and $0.6$. 
This suggests that  the effect of the correction to the 
hyperscaling relation due to the 
large mass deformation appears also in the case of 
finite-size hyperscaling relation. Note that data we used 
here are in the range of $m_P/\Lambda \gtrsim 0.3$. In that range, from 
Fig.~\ref{fig:gamma-eff}, we see that the effective 
anomalous 
dimension takes the value $\gamma_m^{\rm eff} = 0.5 \sim 0.6$. 
This is the reason why the data show good finite-size  
hyperscaling with input value of $\gamma_m^{\rm eff}=
\gamma = 0.5 \sim 0.6$.
We have done the same analysis for SU(3) gauge theory 
with $N_f=14$ and $16$, and found similar results.

It is interesting to ask whether there is a function which can 
be fitted to all the data shown in Fig.~\ref{fig:Nf12}. 
We tried the following form of fit function, and found that 
it can be globally fitted to all the data fairly well:
\beq
 m_0 (L, m_P) = 
 \left[  
 A m_P (1+B m_P^{2-2\gamma})^{1/(1+\gamma)}
 + \frac{C}{L}
 \right]^{1+\gamma}.
\label{eq:fitting}
\eeq
Here, $A, B, C$ and $\gamma$ are fit parameters, and 
it is understood that all the dimensionful 
quantities are normalized by $\Lambda$.
This fit function is similar to the form of finite-size  
hyperscaling relatinon in Eq.~(\ref{eq:HS2}): 
When one takes $B=0$, $L m_P$ is expressed by a 
function of $x= L m_0^{1/(1+\gamma)}$. 
The term proportional to $B$ represents the effect of 
mass correction to the hyperscaling relation.
The best-fit values we obtained for fit parameters are:
$A=1.52, B=-0.512, C=0.323$ and $\gamma = 0.794$.
It is remarkable that we obtained value of $\gamma$ 
which is quite close to the value of $\gamma_m^\ast=0.8$.
For comparison, we also did fitting with fixing $B=0$, 
and found that the best-fit value of $\gamma=0.52$. 
This is consistent with Fig.~\ref{fig:Nf12HS}, 
in which it was shown that finite-size scaling (without 
correction term) is approximately 
satisfied when $\gamma = 0.5\sim 0.6$. 
Of course, the power of the second term in the RHS of 
Eq.~(\ref{eq:fitting}), namely $2-2\gamma$, is specific to the ladder SD analysis, 
though it is worth trying to do fitting lattice data with using 
the above fit function. One could also make the power of 
the second term in the RHS of 
Eq.~(\ref{eq:fitting}) as free parameter. If the chi-square 
of the fitting significantly reduces by the inclusion of the 
correction term, or even if the chi-square does not change 
very much but the value of $\gamma$ significantly changes, 
it is very possible that the best-fit value of $\gamma$ which 
is obtained with correction term into consideration is close 
to the actual value of $\gamma_m^\ast$.

\subsection{Violation of the hyperscaling relation in theories with 
spontaneous chiral symmetry breaking}

Here, by the same procedure used in the previous 
subsection, we study the finite-size hyperscaling relation 
in theories with spontaneous chiral symmetry breaking. 
In the case of theories with spontaneous chiral symmetry breaking, 
mass gap exists even in the chiral limit, and therefore 
an IRFP is only approximate. Here, we show two examples: 
one is SU(3) gauge theory with $N_f=9$, and the other is that 
with $N_f=11$. The former is an example of a theory which is 
far away from the conformal window, in which the infrared 
conformality is expected to be largely violated. The latter is 
an example of a theory which resides close to the conformal 
window, and the breaking of the infrared conformality due to 
the spontaneous chiral symmetry breaking is expected to be 
small.

\begin{figure}
  \begin{center}
    \includegraphics[scale=0.4]{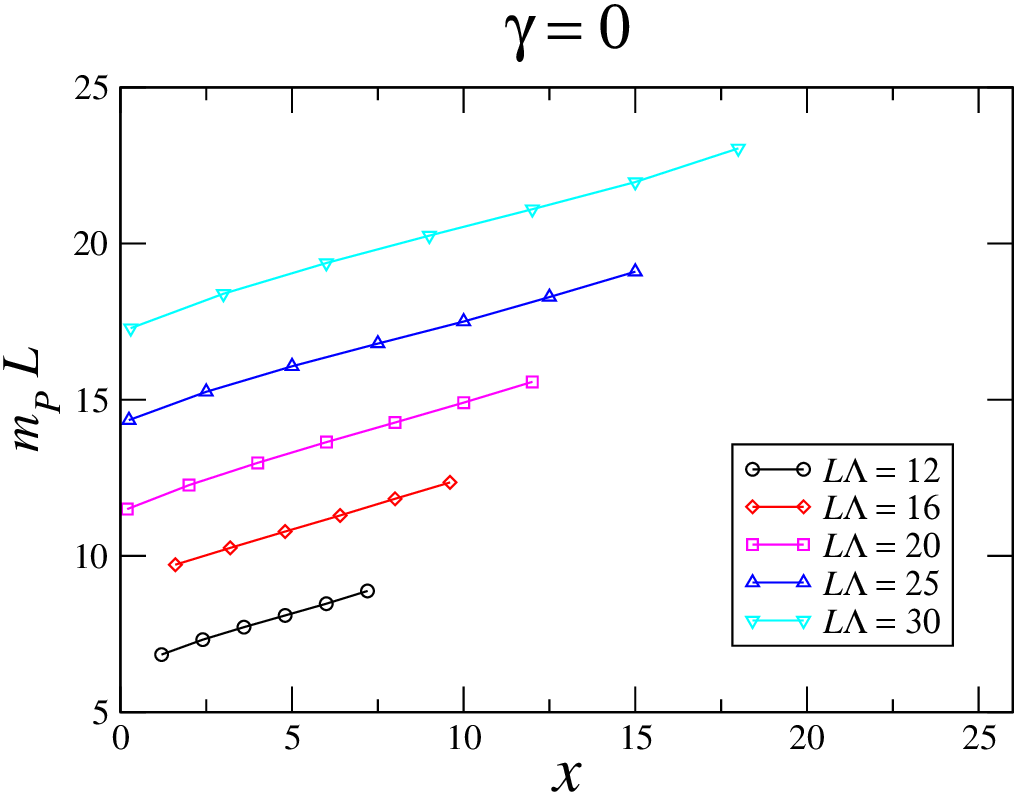}\ \ \ \ \ \ \ \ \ 
    \includegraphics[scale=0.4]{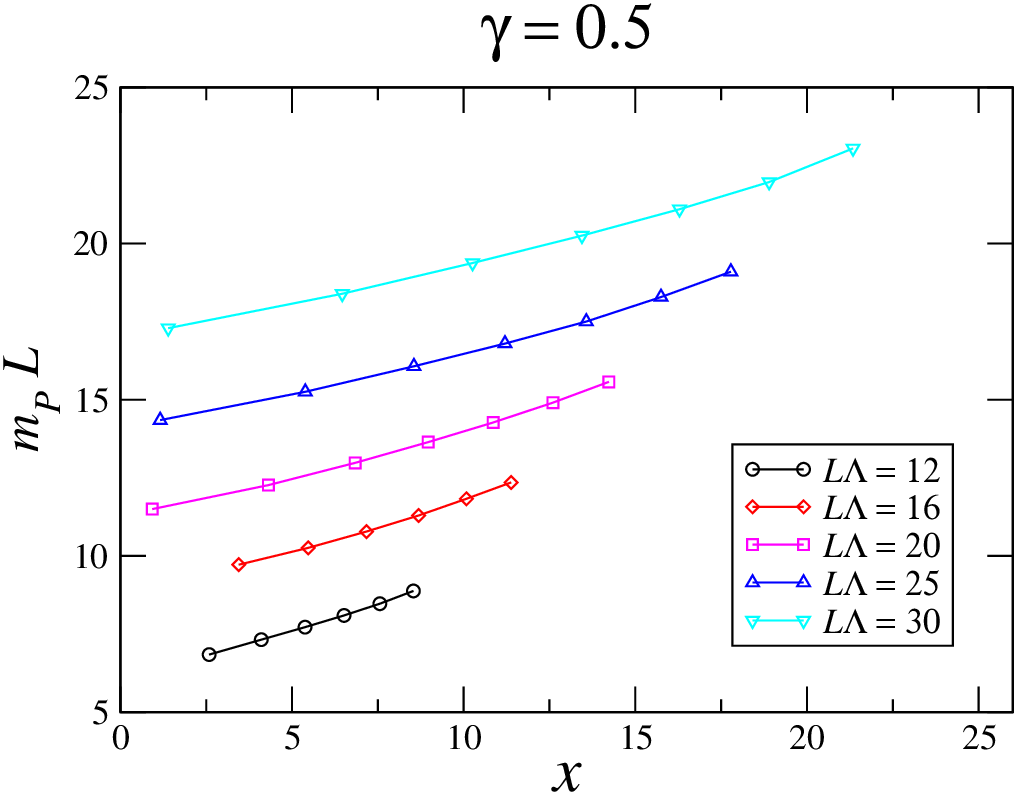}\\
 \vspace*{10mm}
    \includegraphics[scale=0.4]{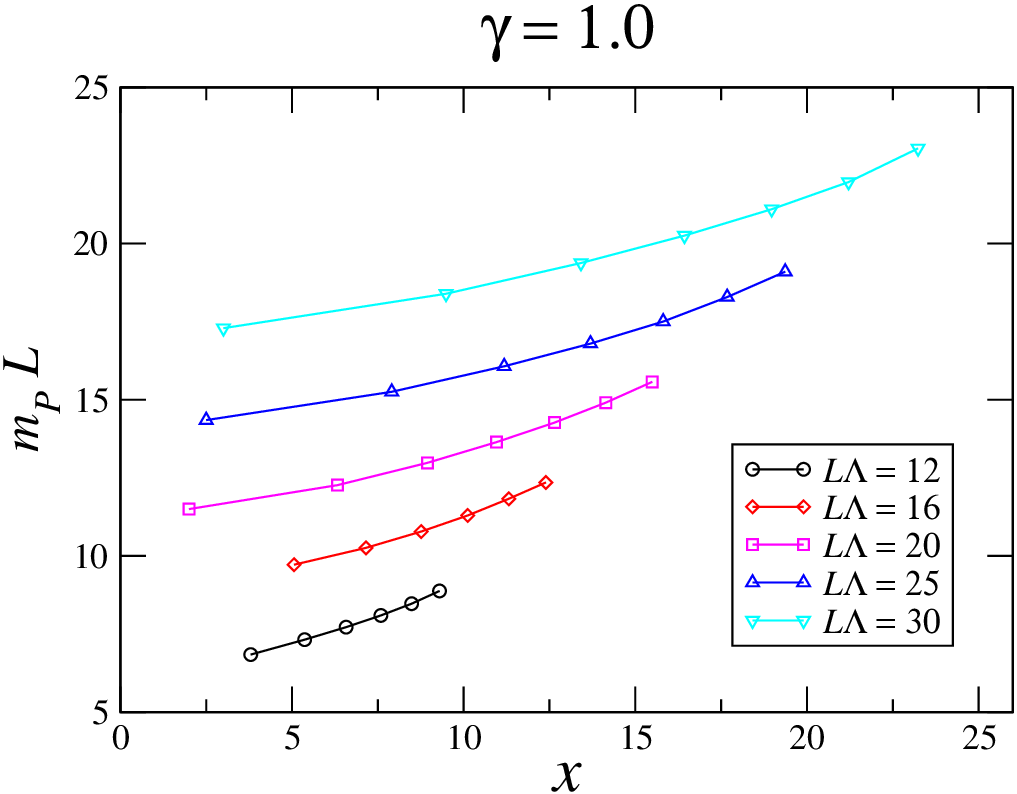}\ \ \ \ \ \ \ \ \ 
    \includegraphics[scale=0.4]{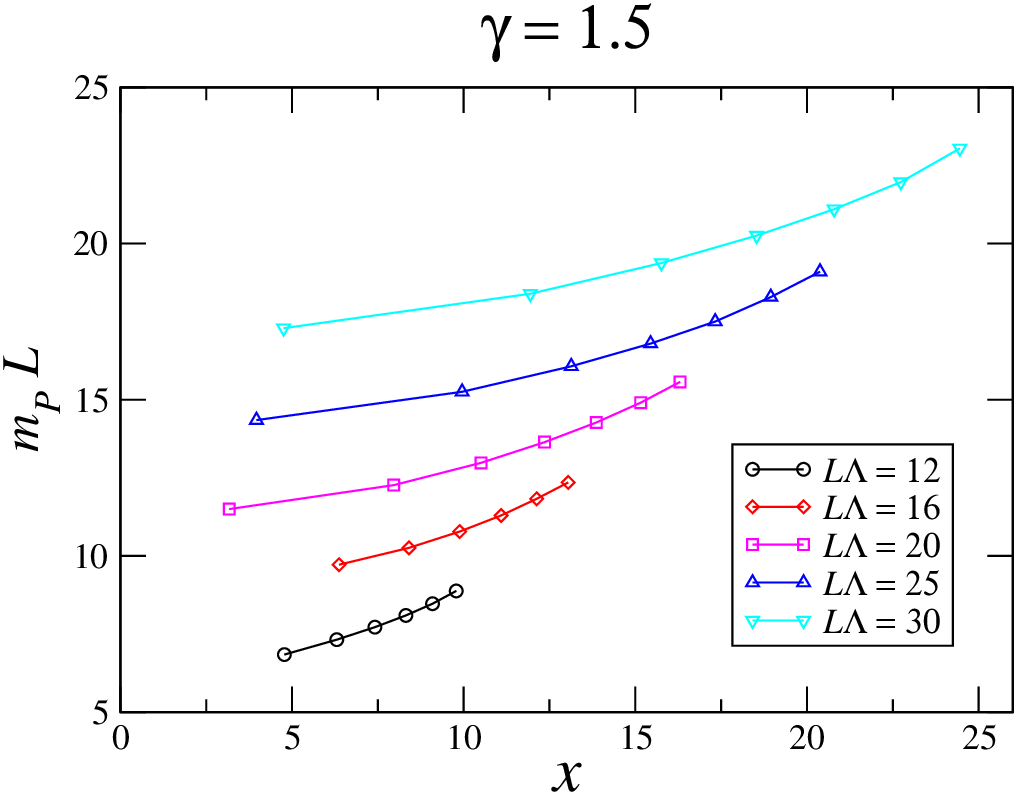}\\
 \vspace*{10mm}
    \includegraphics[scale=0.4]{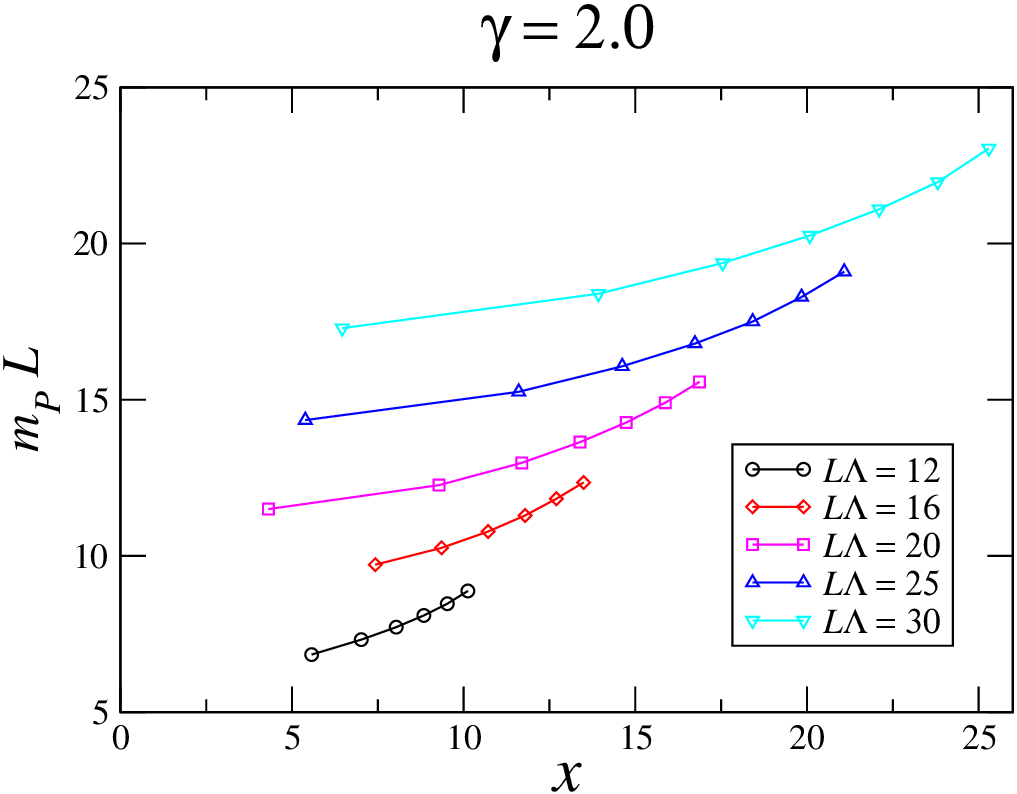}
  \end{center}
\caption{
Values of $m_P L$ obtained from the finite-volume SD equation 
as a function of 
$x\equiv L\Lambda (m_0/\Lambda)^{1/(1+\gamma)}$ for 
$\gamma = 0, 0.5, 1.0, 1.5$ and $2.0$ in 
SU(3) gauge theory with 9 fundamental fermions.
Data for $L\Lambda =12, 16, 20, 25$ and $30$ are plotted 
as different symbols.
}
\label{fig:Nf09HS}
\end{figure}

In Fig.~\ref{fig:Nf09HS}, we show the plots of 
$m_P L$ obtained from the finite-volume SD equation 
in SU(3) gauge theory with 9 fundamental fermions 
as a function of 
$x\equiv L\Lambda (m_0/\Lambda)^{1/(1+\gamma)}$ for 
$\gamma = 0,\, 0.5,\, 1.0,\, 1.5$ and $2.0$.
Data for $L\Lambda =12, 16, 20, 25$ and $30$ are plotted 
as different symbols.
As we expected, since the infrared conformality is largely broken 
due to the spontaneous chiral symmetry breaking, large violation 
of hyperscaling relation is observed. Note that  the dynamically generated mass for $N_f=9$  is $m_D/\Lambda \simeq 0.58$, where $m_D$ is the value of $m_P$ obtained by the spontaneously broken solution of the ladder SD equation in the chiral limit $m_0\equiv 0$. ($m_D$ is estimated by the SD equations, Eq.~(\ref{SD2}), 
in the chiral limit $m_0 = 0$.)
This is compared with the typical values in Fig.~\ref{fig:Nf09HS}:  $m_P/\Lambda=0.58-0.77$ for $L\Lambda=30$.

\begin{figure}[h]
  \begin{center}
    \includegraphics[scale=0.65]{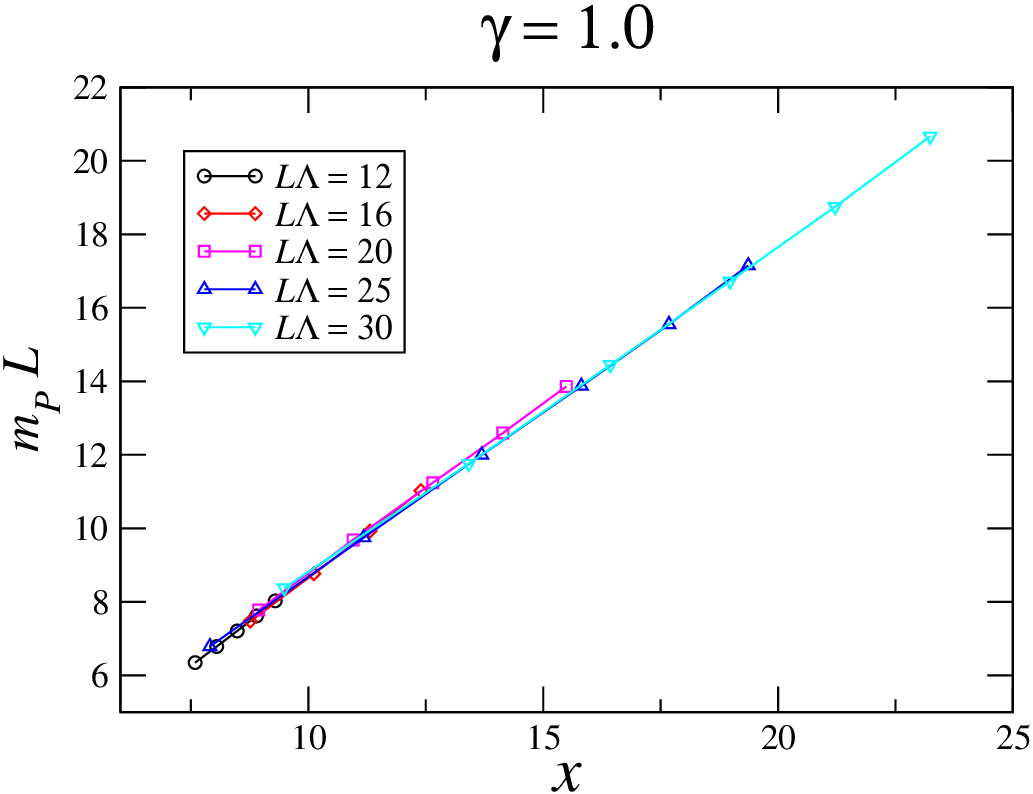}
  \end{center}
\caption{
Values of $m_P L$ obtained from the finite-volume SD equation 
as a function of 
$x\equiv L\Lambda (m_0/\Lambda)^{1/(1+\gamma)}$ for 
$\gamma = 1.0$ in SU(3) gauge theory with 11 fundamental fermions.
Data for $L\Lambda =12, 16, 20, 25$ and $30$ are plotted 
as different symbols.
}
\label{fig:Nf11HS}
\end{figure}

On the other hand, a 
similar plot for SU(3) gauge theory 
with 11 fundamental fermions
is given Fig.~\ref{fig:Nf11HS}. 
We show the result for $\gamma=1.0$, 
with which we found data are best aligned each other. 
Again, as we 
expected, since the theory is close to the chiral restoration point, 
and the effect of the spontaneous 
chiral symmetry breaking is small, the violation of hyperscaling relation 
is small.  Note that $m_D/\Lambda \simeq 0.05$ for $N_f=11$,  while typical values of  $m_P$ in Fig.~\ref{fig:Nf11HS}
are $m_P/\Lambda 
= 0.28-0.69 \, (\gg m_D/\Lambda)$ for $L\Lambda=30$.
Of course, one can see that there is a small amount of 
misalignment. 
However, let us imagine those were data obtained from lattice 
simulations, and each data point has, say, a few percent error bar, 
in which case, the data might look consistent with conformal 
hyperscaling.
Therefore, when one obtained data which 
look consistent with conformal hyperscaling with a large mass 
anomalous dimension, there is 
a possibility that the theory is exactly the one the technicolor 
model favors, namely the dynamics with spontaneous 
chiral symmetry breaking at hierarchically small scale compared 
to $\Lambda$ with large anomalous dimension.

\section{Summary and Discussion}
In this paper, we studied corrections to 
the conformal hyperscaling relation by taking 
the example of SU(3) gauge theories with various 
number of fundamental fermion. 
From the analytical expression of the solution of the 
ladder SD equation, we identified the form of the 
leading correction to the hyperscaling relation.
We found that the anomalous dimension, when 
identified through the hyperscaling relation neglecting 
these corrections (which we denoted as $\gamma_m^{\rm eff}$), 
tends to be lower than the real value at the fixed point. 

We further studied finite-size hyperscaling relation 
through the ladder SD equation in a finite space-time with 
the periodic boundary condition. We found that 
the anomalous dimension, when identified through the finite-size hyperscaling relation neglecting the mass corrections 
as is often done in the lattice analyses,  yields almost the same value 
as that in the case of the infinite space-time neglecting the mass 
correction, i.e.,  a  lower value than $\gamma_m^*$.
The introduction of the finite size of space-time should also break 
the infrared conformality, though we found that the correction to 
the hyperscaling relation due to 
the finite-volume effect seems to be negligible at least  
in the range of $L$ we studied in this paper. This can be seen 
from the fact that the finite-size hyperscaling relation is 
approximately satisfied with $\gamma_m^{\rm eff}$ which 
is obtained from the infinite-volume analysis. If $1/L$ 
correction were large, there must have been a visible violation 
of hyperscaling relation caused by it. The smallness of 
correction coming from finite-size effect can also be understood from 
the fact that a function with a form shown in Eq.~(\ref{eq:fitting}), 
in which only mass correction is taken into account, 
can be fitted to all the data in Fig.~\ref{fig:Nf12} pretty well. 

We also applied the 
finite-volume SD equation to the
chiral-symmetry-breaking phase and found that 
when the theory is close to the critical point such that  the dynamically generated mass is much smaller than 
the explicit breaking mass, the finite-size hyperscaling relation is still operative,  with the mass corrections to the
anomalous dimension being somewhat involved, however. 

From a lattice simulation point of view, 
there are several things we can learn from the results of 
the present paper. 
When the input bare mass is not small enough, and data 
are not precise enough to find the mass correction,  
finite-size hyperscaling plot might give fairly good aligned picture 
with a value of the mass anomalous dimension which is much 
smaller than the value at the IRFP. If data are precise 
enough, one could notice misalignment of data which is 
caused by the fact that the value of $\gamma_m^{\rm eff}$ is 
different for different values of the meson mass $M$. However, if one  
didn't know that the misalignment is fake coming from the 
correction term, one could draw conclusion that 
the theory is not infrared conformal, even though it actually is. 
As we mentioned at the end of the previous section, opposite 
could also happen, namely, even if a theory  
is actually in the chiral symmetry breaking phase, one could 
draw conclusion that it is infrared conformal especially when 
the amount of 
spontaneous chiral symmetry breaking is very small and/or 
data is not precise enough. 
Thus, careful attitude is important when one judges whether a 
theory is infrared conformal or not by using hyperscaling analysis. 
However, main message of our analysis is that if one observed a certain type of the finite-size hyperscaling 
relation (with some finite mass corrections), it already hints the remnant of the IR conformal theory no matter it may be  in the broken phase (applicable to the walking technicolor) 
or the conformal window: It implies a new situation of the 4-dimensional non-Supersymmetric 
gauge theories and a new phenomenological application.

In this paper, we studied large $N_f$ QCD as a concrete example 
for the study of hyperscaling relation, 
though the extension to 
different number of color and different fermion representation is 
straightforward. This is because, in the context of the SD equation 
with the improved ladder approximation, $C_2\, \bar{g}^2$ appearing 
in the equation is the 
only quantity which differentiate different theories, and with the 
simplification of the running coupling adopted in the 
current study, this is proportional to 
$\alpha_\ast/\alpha_{\rm cr}$. Therefore the only relevant 
thing is how close the value of the running coupling at 
the IRFP is to the critical coupling.

It is also interesting to ask what is the best way 
of analyzing data to extract the correct picture. The SD equation 
gives us a nice playground to try to find an analysis method which 
works well for finding correct picture of a given theory 
since it can generate as many data as we like, 
and we know the ``answer", namely, whether the theory possesses 
an IRFP, and also the value of the mass anomalous dimension 
in that theory. 
We can try several different 
analysis methods with those generated data, and compare the 
results with the answer. By doing so, we can tell which analysis 
method 
produces the answer rather correctly. We tried fitting using the 
fit function shown in Eq.~(\ref{eq:fitting}) as an example of such 
studies, and found that it works quite well extracting the 
true value of $\gamma_m^\ast$ of the theory. 
Of course, it is worth investigating further in this direction. 
Various different analysis methods should be studied 
for the purpose of finding a practical method which can 
extract a more correct picture from lattice data. 

\acknowledgments
We thank Luigi Del Debbio and 
Julius Kuti for fruitful discussions during
their  stay at the Kobayashi-Maskawa Institute. 
We also thank George T. Fleming and Katsuya Hasebe 
for valuable discussions. This work was supported by 
the JSPS Grant-in-Aid for Scientific Research 
(S) \#22224003,  
(C) \#23540300 (K.Y.), 
(C) \#21540289 (Y.A.), 
and also by 
Grants-in-Aid of the Japanese Ministry for Scientific Research on Innovative 
Areas \#23105708(T.Y.).  


%

\end{document}